\documentclass[fleqn,usenatbib]{mnras}

\usepackage{newtxtext,newtxmath,ulem}
\usepackage{tikz}
\usetikzlibrary{decorations.pathreplacing, calc}

\tikzstyle{eq} = [rectangle, text width=27em, rounded corners, minimum height=3em] 
   
\tikzstyle{line} = [draw]
\tikzstyle{data} = [rectangle, draw, fill=white!10,
    text width=3em, text centered, rounded corners, node distance=1.5cm, inner sep=4pt]     
\tikzstyle{eq} = [rectangle, text width=27em, rounded corners, minimum height=3em] 
   
\tikzstyle{prod} = [draw, node distance=1cm,
    minimum height=2em]
\tikzstyle{prob} = [rectangle, draw, fill=green!10,
    text width=7em, text centered, rounded corners, node distance=1cm, inner sep=4pt]
\tikzstyle{det} = [rectangle, draw, fill=blue!10,
    text width=7em, text centered, rounded corners, node distance=1cm, inner sep=4pt] 
    
\tikzstyle{label} = [rectangle, 
    text width=8em, text centered, rounded corners, node distance=2cm, inner sep=4pt]    

\usepackage[T1]{fontenc}

\DeclareRobustCommand{\VAN}[3]{#2}
\let\VANthebibliography\thebibliography
\def\thebibliography{\DeclareRobustCommand{\VAN}[3]{##3}\VANthebibliography}

\usepackage{graphicx}	
\usepackage{amsmath}	

\title[Field-level inference of cosmic shear]{Field-level inference of cosmic shear with intrinsic alignments and baryons}

\author[N.\ Porqueres et al.]{
Natalia Porqueres$^{1}$\thanks{natalia.porqueres@physics.ox.ac.uk},
Alan Heavens$^{2}$,
Daniel Mortlock$^{2,3,4}$,
Guilhem Lavaux$^{5}$ and T.\ Lucas Makinen$^{2}$
\\
$^{1}$Department of Physics, University of Oxford, Denys Wilkinson Building, Keble Road, Oxford OX1 3RH, UK\\
$^{2}$Imperial Centre for Inference and Cosmology (ICIC) \& Astrophysics group, Department of Physics, Imperial College, Blackett Laboratory, \\ Prince Consort Road, London SW7 2AZ, UK\\
$^{3}$Department of Mathematics, Imperial College London, London, SW7 2AZ, UK\\
$^{4}$The Oskar Klein Centre, Department of Astronomy, Stockholm University, Albanova, SE-10691 Stockholm, Sweden\\
$^{5}$CNRS \& Sorbonne Universit\'{e}, UMR7095, Institut d'Astrophysique de Paris, F-75014, Paris, France
}

\date{Accepted . Received ; in original form }

\pubyear{2021}

\begin{document}
\label{firstpage}
\pagerange{\pageref{firstpage}--\pageref{lastpage}}
\maketitle

\begin{abstract}
     \noindent
          We construct a field-based Bayesian Hierarchical Model for cosmic shear that includes, for the first time, the important astrophysical systematics of intrinsic alignments and baryon feedback, in addition to a gravity model. We add to the BORG-WL framework the tidal alignment and tidal torquing model (TATT) for intrinsic alignments and compare them with the non-linear alignment (NLA) model. With synthetic data, we have shown that adding intrinsic alignments and sampling the TATT parameters does not reduce the constraining power of the method and the field-based approach lifts the weak lensing degeneracy. We add baryon effects at the field level using the enthalpy gradient descent (EGD) model. This model displaces the dark matter particles without knowing whether they belong to a halo and allows for self-calibration of the model parameters, which are inferred from the data. We have also illustrated the effects of model misspecification for the baryons.  The resulting model now contains the most important physical effects and is suitable for application to data.
\end{abstract}

\begin{keywords}
    cosmology:large-scale structure of Universe -- methods:data analysis -- weak gravitational lensing
\end{keywords}



\section{Introduction}

\begin{figure*}
    \includegraphics[width=0.85\hsize]{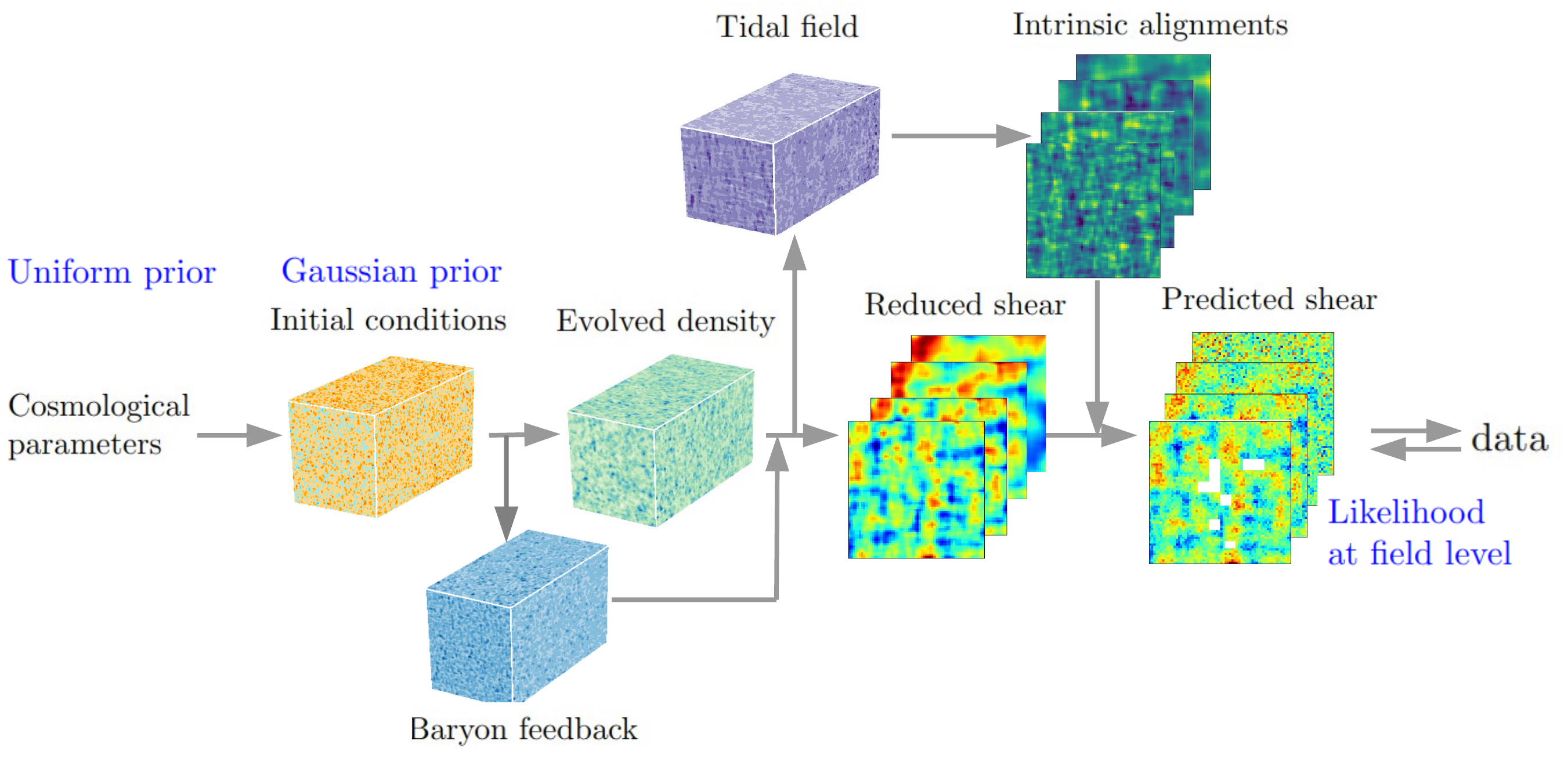}
    \caption{Representation of the forward model. The text in blue indicates the probability distributions.}
    \label{fig:borg_diagram}
\end{figure*}

Weak gravitational lensing is a powerful probe of cosmology as it is sensitive to the growth of structures and the geometry of the Universe. Cosmic shear analyses based on the two-point statistics have provided constraints on the cosmological parameters \citep{Troxel2018, Hikage19, Hamana2020, Asgari2021, Amon2021, Secco2021}. However, the two-point summary statistics are sub-optimal for non-Gaussian fields and discard information, typically resulting in a degeneracy in the posterior of the main parameters that we can measure with weak lensing: $(\Omega_m, \sigma_8)$, which describe the amount of matter in the Universe and how clustered this is. For this reason, several alternative data analysis techniques have been developed, such as peak count statistics \citep{Jain2000, Dietrich2010, Maturi2011, Lin2015, Liu2015a, Kacprzak2016Peaks, Petri2013, Peel2017, Fluri2018Peaks, Martinet2018Peaks, Shan2018Peaks, Harnois2021Peaks, Zurcher2022, Liu2023}, the probability distribution function \citep{Boyle2021PDF, Martinet2021PDF, Boyle2022}, shear clipping \citep{Giblin2018}, and machine learning approaches \citep{Gupta2018ML, Fluri2018ML, Jeffrey2021, Ribli2019ML, Fluri2022}. \cite{Euclid2023Comparison} presented a comparison between many of these methods, demonstrating their potential to reduce the marginal uncertainties on $\Omega_m$ and $\sigma_8$ from those of a two-point correlation function. However, these methods require assumptions on the sampling distribution of the summary statistics and a covariance matrix, which is difficult to compute accurately. An alternative to summary statistics is incorporating the data into a forward model through a data assimilation approach. Several forward modelling approaches have been developed for lensing \citep{Alsing16, Boehm17, Alsing17, BORG-WL, Fiedorowicz2021, BORG-WL2022, Boruah2022, 2Fiedorowicz022, Remy2022, Loureiro2023} and they differ in their assumptions and the quantities they sample.

One such approach is BORG-WL \citep{BORG-WL, BORG-WL2022}, which is based on the Bayesian Reconstruction from Galaxies \citep[BORG, ][]{jasche2013bayesian, BORG-3} and differs from the other forward models in incorporating a physical description of structure formation, which allows us to sample the initial conditions and the cosmological parameters simultaneously. \cite{BORG-WL2022} showed that a field-based analysis can lift the weak lensing degeneracy, yielding marginal uncertainties on $\Omega_m$ and $\sigma_8$ up to a factor 5 smaller than those from a two-point power spectrum analysis on the same simulated data. However, our previous work did not include intrinsic alignments and baryon feedback, which are essential to apply the method to real cosmic shear measurements. 

Tidal processes during galaxy formation generate intrinsic shape correlations \citep{Heavens2000, Croft2000, Catelan2001,  Mandelbaum2006, Hirata2007, Joachimi2011, Blazek2011, Joachimi2011, Blazek2019}. These intrinsic alignments of galaxies are a contaminant of weak lensing measurements, which assume that galaxy shapes are uncorrelated. Since these intrinsic alignments can lead to biases \citep{Troxel2015, Samuroff2019, Blazek2019}, it is necessary to include them in the data model. There are several models to describe the intrinsic alignments, including the non-linear tidal alignment \citep[NLA, ][]{Bridle2007}, the tidal torquing \citep{Hirata2004, Catelan2001} and a combination of both: the tidal alignment and tidal torquing model \citep[TATT,][]{Blazek2019}. These models have also been studied in field-level approaches \citep{Harnois2021, Tsaprazi2022, Kacprzak2023}, peak counts \citep{Davies2022, Zhang2022, Aucoberry2022, Burger2023, Liu2023} and persistent homology \citep{Heydenreich2022}. However, there is still a large uncertainty regarding the strength of intrinsic alignments. DES-Y1 \citep{Troxel2018}, KiDS \citep{Asgari2021} and HSC \citep{Hikage19, Hamana2020} used the NLA model and reported non-zero values for the NLA parameters, while DES-Y3 \citep{Secco2021} used both NLA and TATT. Some recent studies \citep{Blazek2015, Fortuna2020, Troxel2018} found that the NLA model is disfavoured over more complex models, while \cite{Secco2021} found that TATT and NLA are consistent, but TATT is unnecessarily flexible for the analysis of DES-Y3 data and degrades the cosmology constraints.

Baryon feedback suppresses structure formation at small scales and also affects weak lensing surveys. Gravity leads to collapse, but baryons resist due to the gas pressure. In addition, baryon feedback can transport large amounts of gas to the outskirts of halos, which leads to an expansion of the dark matter halos and reduction of their mass \citep{Duffy2010, Mccarthy2011, Teyssier2011, Velliscig2014}. Many of the descriptions of baryonic effects are halo-based and assume that the baryons only affect the matter distribution within halos \citep{Rudd2008, Semboloni2011, Mohammed2014, Velliscig2014, Mead2015, Copeland2018, Mead2021, Peacock2000, Seljak2000, vanDaleen2019}. At the field level, several models follow the same approach and displace the dark matter particles inside the halos to mimic hydrodynamical simulations \citep{Schneider2015, Schneider2019, Arico2021, Lu2022, LeePeakBaryons2023}.  However, \cite{Liu2023Baryons} has found that baryons also affect filaments, walls and voids. Therefore, it is important to model the effects of baryons in the whole cosmic web rather than just in the halos. The enthalpy gradient descent method \citep[EGD,][]{Dai2018} and the Lagrangian deep learning method \citep{Dai2021} quantify the baryon effects without relying on halos.

In this work, we extend the forward model of BORG-WL to include the TATT model for intrinsic alignments and the EGD model of baryon feedback. We also extend our framework to sample the parameters of these models. Rather than calibrating the baryon parameters to simulations, we sample them and allow the baryon model to self-calibrate from the data. By including these systematic effects in our pipeline, we bring BORG-WL closer to the real data application, making it the first forward model approach to include baryons and intrinsic alignments.

This paper is organised as follows. Section~\ref{sec:data_model} describes the data model, including intrinsic alignments and baryon feedback. In Section~\ref{sec:method}, we present the sampling methods of BORG-WL. Section~\ref{sec:mock_data} describes the simulated data we used to validate our approach. The results are discussed in Section~\ref{sec:results}, which includes a discussion on model misspecification. Finally, we summarise the results in Section~\ref{sec:conclusions}.

\section{The data model}
\label{sec:data_model}

\begin{figure}
    \includegraphics[width=\columnwidth]{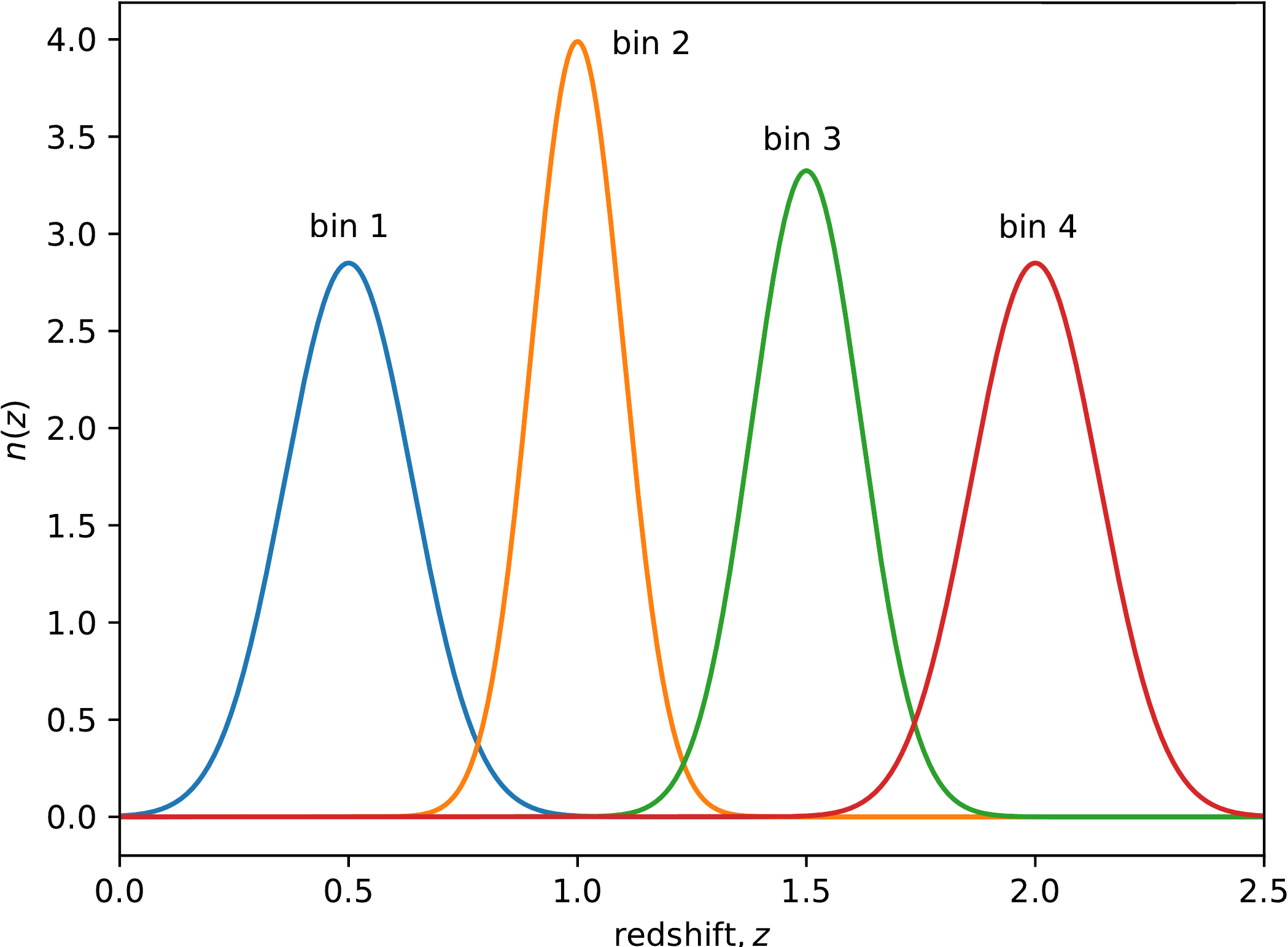}
    \caption{Redshift distributions of sources for each tomographic bin in this analysis. The $n(z)$ are normalised Gaussian distributions.}
    \label{fig:tomobin}
\end{figure}

The effect of weak gravitational lensing on a galaxy can be described by the shear $\gamma$, quantifying the distortion of the galaxy image, and the convergence $\kappa$, which indicates the variation in angular size. However, tidal processes during galaxy formation generate intrinsic shape correlations, which we need to treat separately from the weak lensing effects. Since the forward model has access to the tidal field, we can incorporate intrinsic alignments in BORG-WL at the pixel level using the TATT model \citep{Blazek2019}. We also have improved our description of structure formation by adding baryon feedback with the EGD model \citep{Dai2018}. In this section, we describe the components of the data model in BORG-WL.

\subsection{Reduced shear model}
In the flat-sky approximation, which we assume throughout, the shear and convergence fields are related in Fourier space by
\begin{align}
    \Tilde{\gamma}(\boldsymbol{\ell}) = \frac{(\ell_1 + i \ell_2)^2}{\ell^2} \Tilde{\kappa}(\boldsymbol{\ell}),
\label{eq:shear}
\end{align}
where $\boldsymbol{\ell} = (\ell_1, \ell_2)$ is the angular wave vector, and the tilde indicates a Fourier transformed quantity. In practice, we have the following relation between a quantity $X$ and its Fourier transformed representation $\Tilde{X}$:
\begin{equation}
    \Tilde{X} = \left(\frac{L}{N}\right)^2 \mathbf{F} X,
\end{equation}
with $\mathbf{F}$ being the Discrete Fourier Transform as a matrix operation, with $F_{ab} = \exp(-i \vec{k}_a .\vec{x}_b)$, $L$ is the physical size of the patch in the sky (in radians), and $N$ is the number of pixels in each direction of the sky.

The convergence field is obtained as the integral of the matter overdensity along the line-of-sight,
\begin{equation}
    \kappa(\boldsymbol{\vartheta}) = \frac{3 H_0^2 \Omega_\mathrm{m}}{2 c^2} \int^{r_\mathrm{lim}}_0 \frac{rdr}{a(r)} q(r) \delta(r\boldsymbol{\vartheta}, r),
\label{eq:kappa}
\end{equation}
where $\boldsymbol{\vartheta}$ is the coordinate on the sky, $r$ is the comoving distance, $r_\mathrm{lim}$ is the limiting comoving distance of the galaxy sample, $\delta$ is the dark matter overdensity at a scale factor $a$ and 
\begin{equation}
    q(r) = \int^{r_\mathrm{lim}}_r dr' n(r') \frac{r'-r}{r'},
\end{equation}
with $n(r)$ being the redshift distribution of galaxy sources. We assume a spatially flat universe throughout. In our discrete implementation  and using the Born approximation, the radial line-of-sight integral in Equation~\ref{eq:kappa} is approximated by a sum over voxels as
\begin{equation}
    \kappa^b_{mn} =\frac{3 H_0^2 \Omega_\mathrm{m}}{2 c^2} \sum\limits_{j=0}^{N}  \delta_{mnj} \left[\sum\limits_{s=j}^{N} \frac{(r_s - r_j)}{r_s} n^b(r_s) \Delta r_s \right] \frac{r_j \Delta r_j}{a_j},
\end{equation}
where the index $b$ indicates the tomographic bin, and the sub-indices $m$ and $n$ label the pixel on the sky, which is chosen to be large enough to contain many sources. The index $j$ labels the voxels along the line-of-sight at a comoving distance $r_j$. $N$ is the total number of voxels along the line of sight found using a ray tracer. $\Delta r_j$ is the length of the line of sight segment inside the voxel $j$, and $\delta_f$ is the three-dimensional dark matter distribution. The comoving radial distance $r_s$ indicates the distance to the source. The redshift distribution of sources is given by $n^b(z_s)$ for each tomographic bin. Having evaluated $\kappa_{mn}^b$ in this way, we then transform and use Equation~\ref{eq:shear} to obtain the predicted shear field. 

In this work, we do not include the effects of uncertainty in the redshift distributions of the sources. These effects can be included in the inference as associated nuisance parameters \citep[][Kyriacou, in prep.]{Tsaprazi2023}. 

\begin{figure*}
        \centering
        \begin{tabular}{c c}
            \includegraphics[width=0.45\hsize,clip=true]{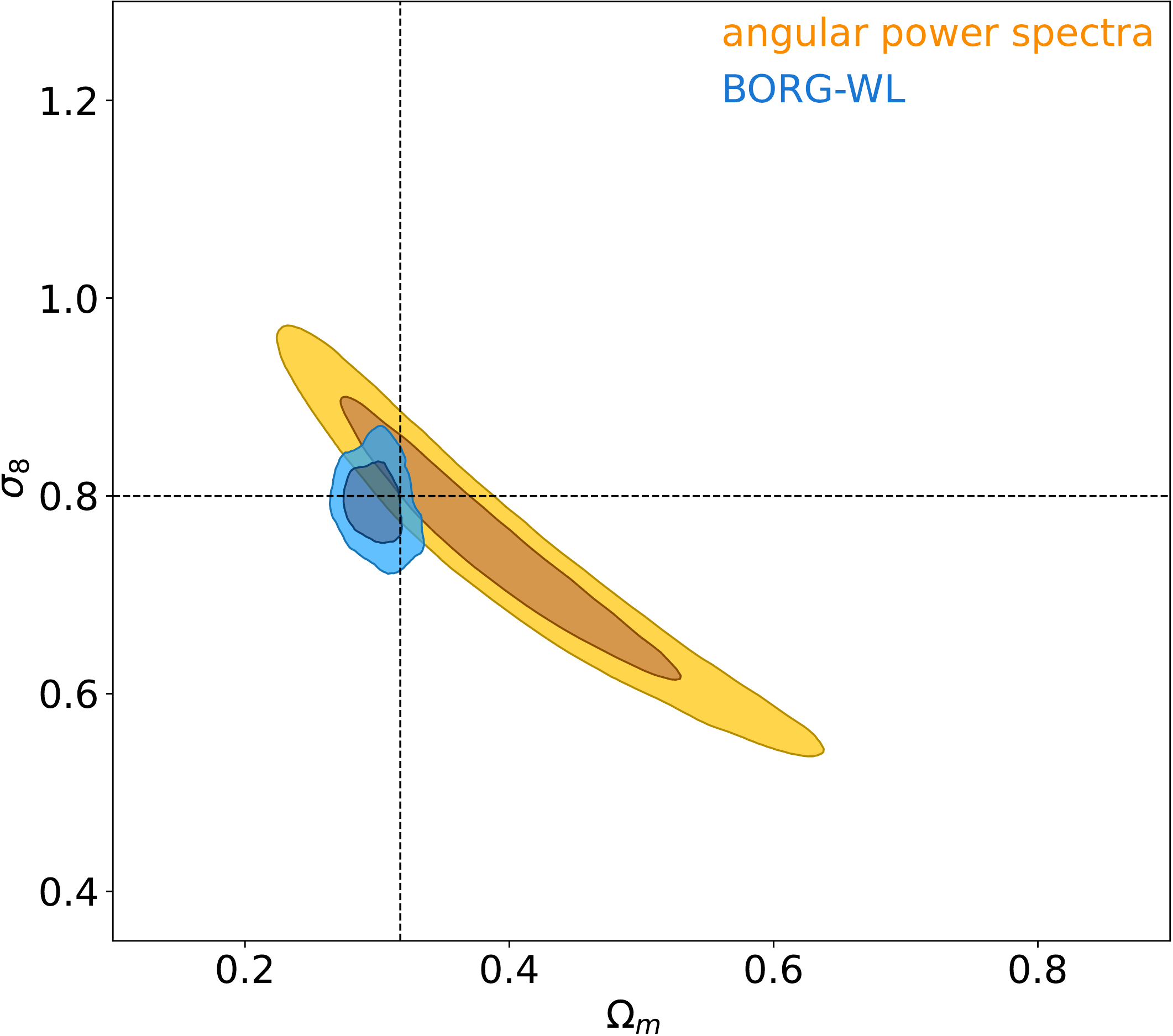} & \includegraphics[width=0.45\hsize,clip=true]{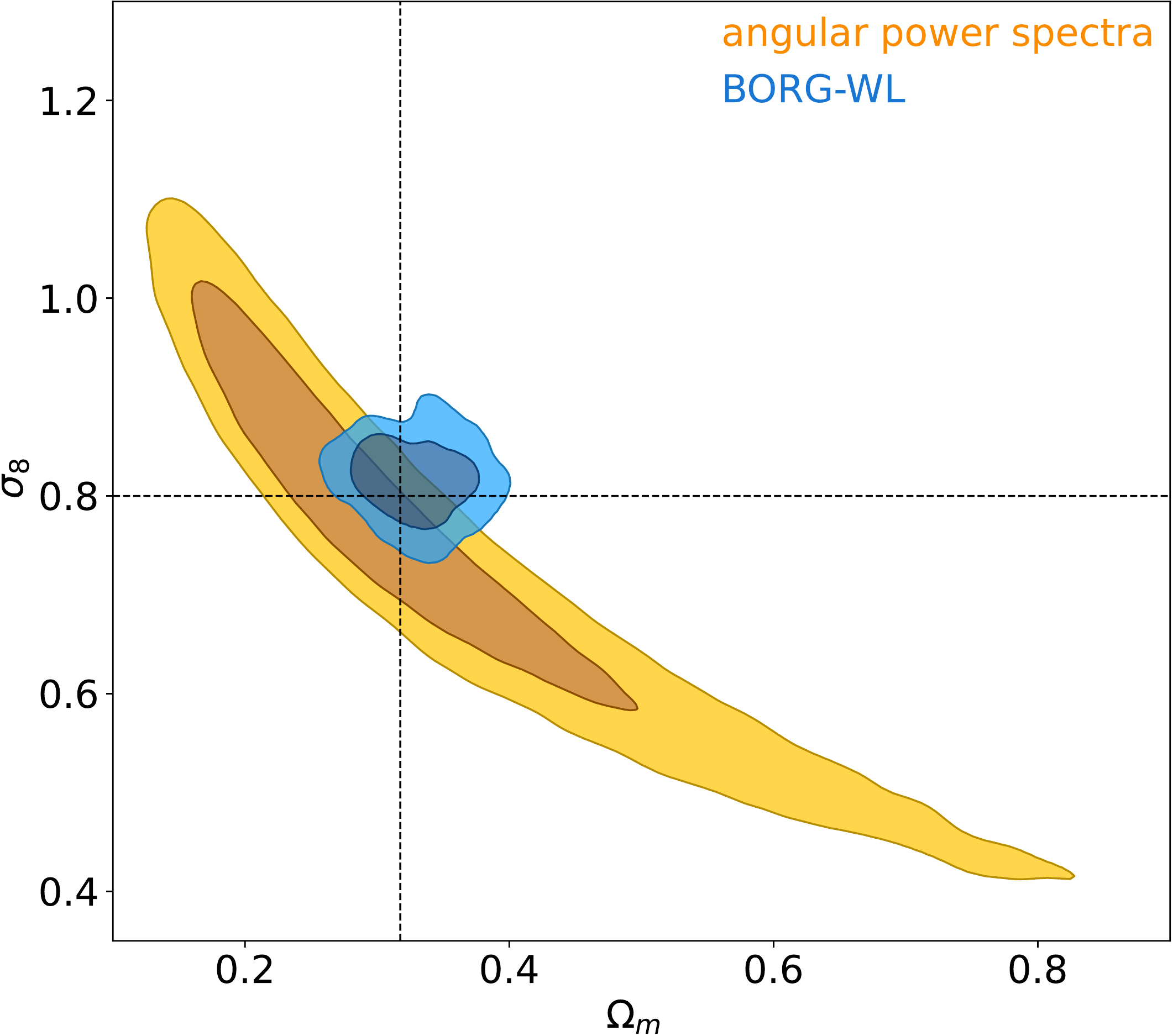} \leavevmode \\
            data: no IA, analysis: no IA & data: TATT, analysis: TATT  
        \end{tabular}    
        \caption{Comparison of $\Omega_\mathrm{m}$-$\sigma_8$ constraints from our method BORG-WL without (left) and with (right) intrinsic alignments. The contours show the 68.3\% and 95.4\%  highest posterior density credible regions. Both runs are self-consistent, meaning the synthetic target data are different for each run.}
    \label{fig:TATT_contours}
\end{figure*}

\begin{figure*}
    \includegraphics[width=\hsize]{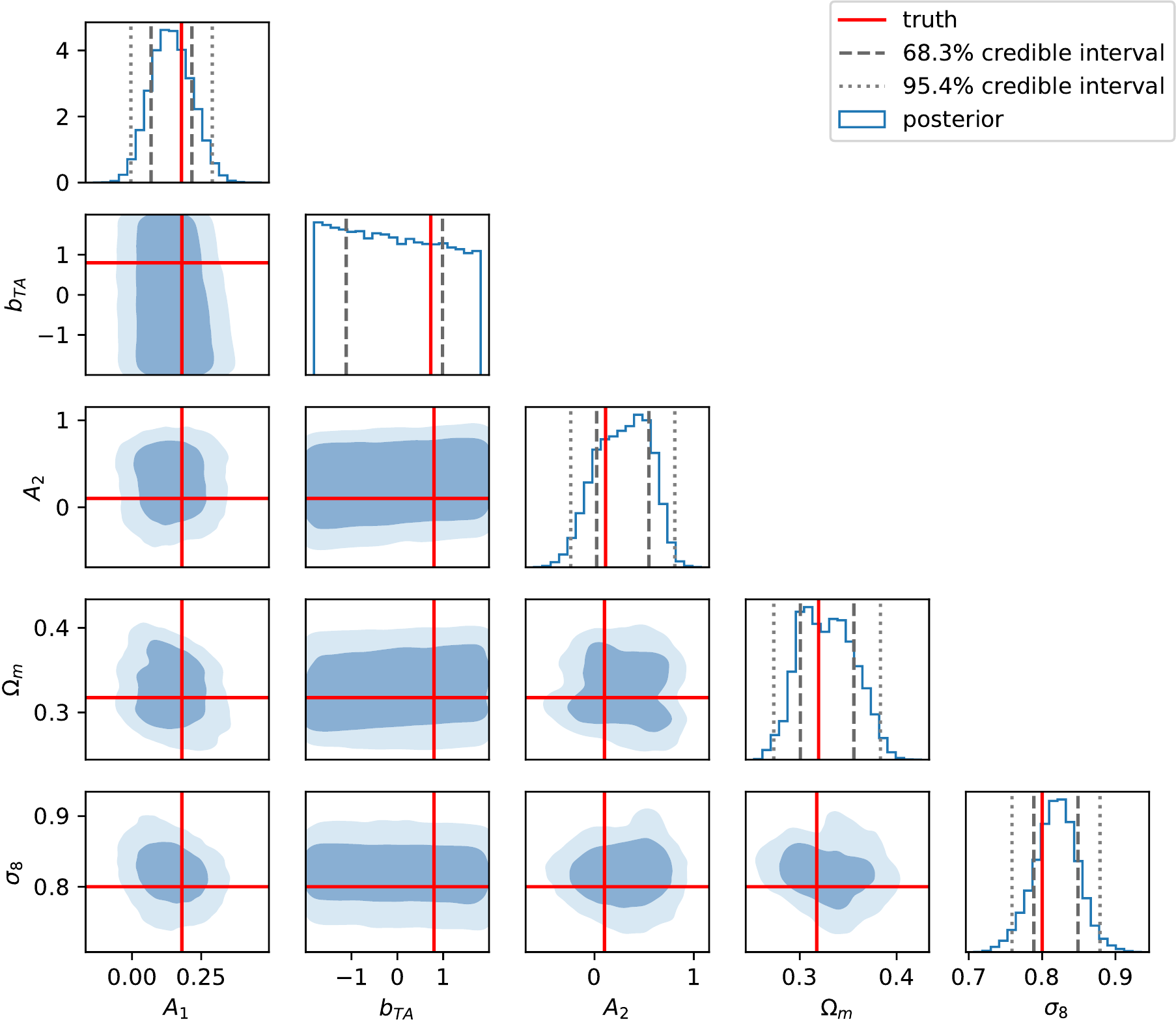}
    \caption{Posterior distribution of the cosmology and TATT parameters inferred from the simulated Dataset 1. The red lines indicate the truth, the grey dashed lines show the 68.3\% credible interval, and the dotted grey lines indicate the 95.4\% credible interval. All the posterior distributions are narrower than the size of the prior except  for $b_\mathrm{TA}$, which closely corresponds to the prior.}
    \label{fig:corner_TATT}
\end{figure*}

\subsection{Intrinsic alignments model}
\label{sec:ia_model}

Since BORG-WL has access to the three-dimensional density contrast $\delta$, we can compute the tidal field $s$ in Fourier space as
\begin{align}
    \Tilde{s}_{ij}(\boldsymbol{k}) = \left(\frac{k_j k_j}{k^2}  - \frac{1}{3} \delta_{ij} \right) \Tilde{\delta}(\boldsymbol{k}), \label{eq:tidal_field}
\end{align}
where the indices label the components of the wave-vector $\boldsymbol{k} = (k_x, k_y, k_z)$ and $k^2 = k_x^2 + k_y^2 + k_z^2$. Following \cite{Blazek2019}, we compute the intrinsic alignments from the tidal field in real space using the TATT model, which accounts for linear alignments and tidal torque:
\begin{align}
    \gamma^\mathrm{IA}_1 (\boldsymbol{r}, \boldsymbol{\vartheta}) &= (C_1 + C_{1\delta} \delta) (s_{xx} - s{yy}) + C_2 (s_{xk} s_{xk} - s_{yk} s_{yk})  \label{eq:TATT_model0} \\
    \gamma^\mathrm{IA}_2  (\boldsymbol{r}, \boldsymbol{\vartheta}) &= 2 (C_1 + C_{1\delta} \delta) s_{xy} + 2 C_2 s_{xk} s_{yk},
    \label{eq:TATT_model}
\end{align}
where the coefficients are
\begin{align}
    & C_1 = -A_1 \Bar{C} \frac{\rho_\mathrm{crit} \Omega_m}{D(z)} \\
    & C_{1\delta} = b_\mathrm{TA}  C_1 \\
    & C_2 = 5 A_2 \Bar{C} \frac{\rho_\mathrm{crit} \Omega_m}{D^2(z)},
\end{align}
with $\Bar{C}$ being a normalisation constant fixed at $\Bar{C} = 5\times10^{-14}h^{-2} \mathrm{M}_\odot $~Mpc$^2$ \citep{Brown2002, DesRobustness}, $\rho_\mathrm{crit}$ being the critical density, and $D(z)$ is the linear growth factor. We sample the TATT parameters $A_1$, $b_\mathrm{TA}$ and $A_2$.

From Equation \ref{eq:TATT_model0} and \ref{eq:TATT_model}, we obtain the intrinsic alignments in the three-dimensional Cartesian box. We then average these over the line of sight to compute their projection on the map as
\begin{align}
    \gamma^\mathrm{IA} (\boldsymbol{\vartheta})= \int^{r_\mathrm{lim}}_0 \gamma^\mathrm{IA} (\boldsymbol{r}, \boldsymbol{\vartheta}) n(r) dr,
    \label{eq:projection_IA}
\end{align}
where $n(r)$ is the redshift distribution of sources.

\subsection{Baryon physics}
\label{sec:baryon_model}

We used the EGD model \citep{Dai2018} to include baryon effects in our forward model. EGD is a numerical scheme based on the motion of particles along the gradient direction of a scalar field generated by the existing density field. 

This model assumes that, to first order, the distributions of baryons and dark matter are the same. It also assumes that the equation of state of the baryons follows a power law
\begin{equation}
    T(\delta) = T_0 (1 + \delta_b)^{\gamma-1},
\end{equation}
where $T_0 = 10^4$ K is the characteristic temperature of the intergalactic medium and $\gamma$ is a free parameter to be sampled. 

By introducing some specific enthalpy, \cite{Dai2018} find that the displacement of the baryon particles follows
\begin{align}
    \boldsymbol{\mathrm{S}}_\mathrm{baryons} = - \frac{\beta}{H_0^2} \frac{k_B T_0}{\mu} \frac{\gamma}{\gamma - 1} \nabla \left[ {\boldsymbol{\mathrm{O}}_\mathrm{J} (1 + \delta)}\right]^{\gamma-1},
\label{eq:baryon_displacement}
\end{align}
where $\beta$ is the amplitude, which we sample in this work, $H_0$ is the Hubble parameter to make $\beta$ dimensionless,  $k_B$ is the Boltzmann constant and $\mu$ is the gas atomic mass, which is set to be the hydrogen atomic mass. The smoothing operator $\boldsymbol{\mathrm{O}}_\mathrm{J}$ is a Gaussian kernel with a smoothing scale $r_\mathrm{J}$ that corresponds to the Jeans' scale, which in Fourier space is 
\begin{align}
 \boldsymbol{\Tilde{\mathrm{O}}}_\mathrm{J}(k) = \mathrm{exp} \left[\frac{-(k r_J)^2}{2} \right].
\end{align}
Since $T_0$ and $\mu$ are degenerate with the parameter $\beta$, we only assign them to the correct order of magnitude. The parameter $\gamma$ determines how the displacement depends on the density field and, therefore, varying $\gamma$ allows fitting the halo mass dependence of the AGN feedback. 

We introduce the baryon correction as a post-processing step after each time step of the gravity solver. Rather than applying a uniform pressure to all the particles, this model displaces only a fraction of the particles.  We note that the EGD model moves the particles independently on which halo they belong to or where in the halo they are located. Since it does not require a halo finder, this model is differentiable, which allows us to use it with our Hamiltonian Monte Carlo sampler.

\subsection{The forward model}

All these fields can be interpreted as latent parameters of a Bayesian hierarchical model as represented in Figure~\ref{fig:borg_diagram}. We start by sampling the cosmological parameter $\boldsymbol{\theta}$. Given a cosmology, we generate a random set of initial conditions  $\delta^\mathrm{ic}$ from a Gaussian prior. The covariance matrix of the initial conditions corresponds to the initial matter power spectrum. In this work, we use the prescription \cite{EH98,EH99}, including baryonic effects, but BORG also has CLASS support \citep{CLASS}. These initial conditions then evolve in time using the non-linear gravity model, which describes the evolution of the dark matter density and accounts for light-cone effects. The dark matter field is corrected for baryon feedback and then used to compute the shear and intrinsic alignment fields. We then compute the estimated shape changes from the intrinsic complex ellipticity $\epsilon^\mathrm{s}$, which includes $\gamma^\mathrm{IA}$, \citep{KilbingerReview}
\begin{align}
    \epsilon = \frac{\epsilon^\mathrm{s} + g}{1 + g^*\epsilon^\mathrm{s}},
    \label{eq:epsilon}
\end{align}
where $g=(\gamma_1 + \mathrm{i} \gamma_2)/(1-\kappa)$ is the reduced shear and the denominator in Equation \ref{eq:epsilon} guarantees that no ellipticity exceeds unity.

\begin{figure*}
        \centering
        \begin{tabular}{c c}
            \includegraphics[width=0.45\hsize,clip=true]{TATT.pdf} & \includegraphics[width=0.45\hsize,clip=true]{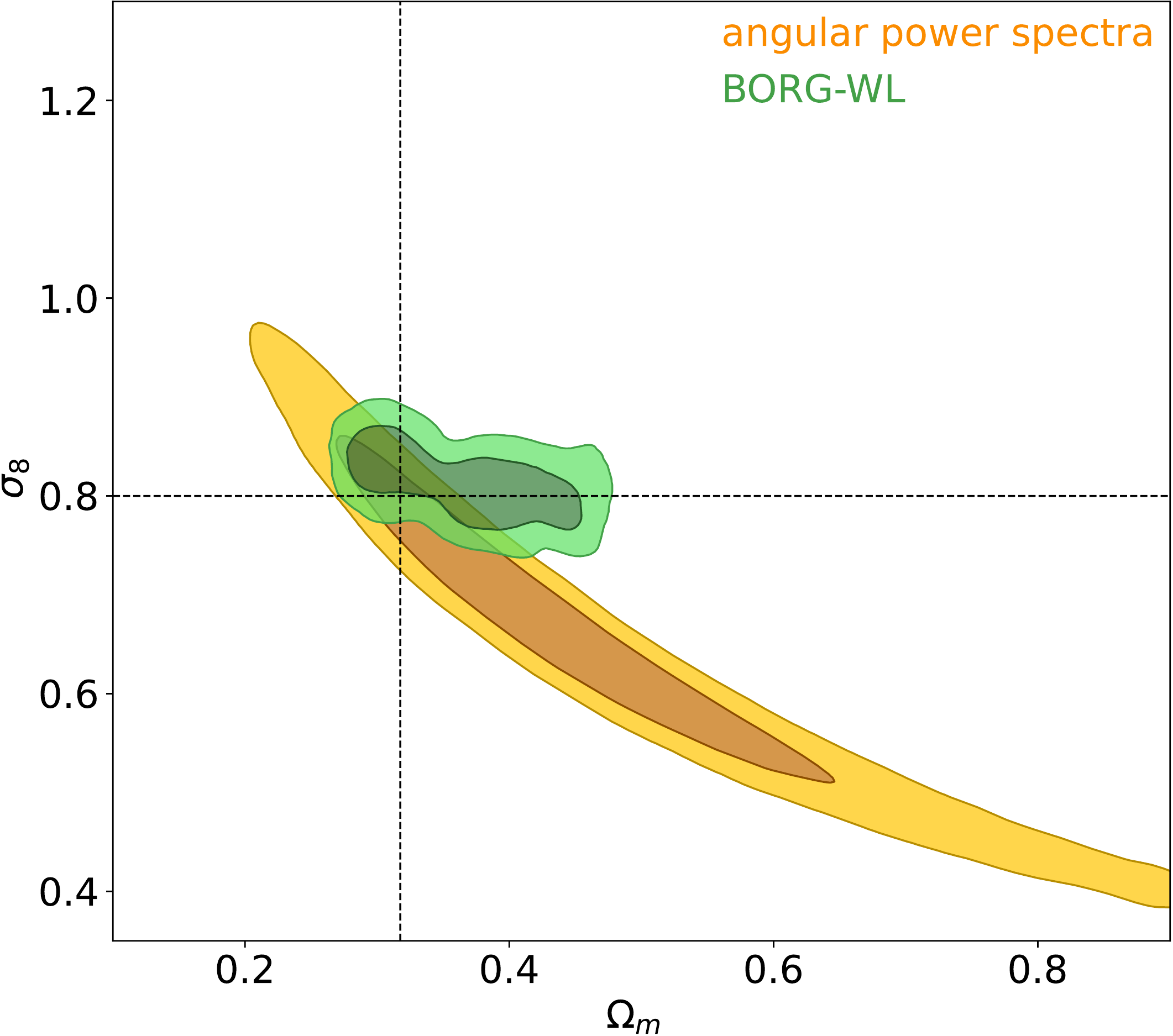} \leavevmode \\
            data: TATT, analysis: TATT & data: TATT, analysis: NLA  
        \end{tabular}
    \caption{Comparison of the $\Omega_\mathrm{m}$-$\sigma_8$ 68.3\% and 95.4\%  highest posterior density credible regions from our method {\sc BORG-WL} from the mock Dataset 1. The different contours correspond to analysing the data with TATT (self-consistent test, left panel) and NLA (right panel). All posteriors are obtained by applying both methods to the same simulated shear data, with 4 tomographic bins and 30 galaxies per square arcmin. The dashed lines indicate the true values of the parameters. The constraints from the angular power spectra use the same models: TATT in the left panel and NLA in the right panel.}
    \label{fig:IA_models}
\end{figure*}

\section{Method} 
\label{sec:method}

Here we briefly describe the BORG-WL method and indicate the changes we have implemented in this work to sample the intrinsic alignments and baryon feedback parameters.

The BORG framework \citep{jasche2010fast, jasche2013bayesian, BORG-3} uses a non-linear gravity model for structure formation. Several options are available based on perturbation theory and particle-mesh simulations \citep{Jasche18BorgPM}. This gravity model connects the initial conditions to the evolved dark matter distribution, allowing us to sample the initial conditions from a Gaussian prior at $a \approx 10^{-3}$. We have modified the gravity model to account for baryon feedback by correcting the positions of the dark matter particles as described in Section~\ref{sec:baryon_model}. This correction is done at each time step: we first obtain the dark matter distribution from the dark matter particles using a cloud-in-cell algorithm and use this density field to compute the displacement field in Equation~\eqref{eq:baryon_displacement}. We then correct the positions of the particles and apply the cloud-in-cell method again to get the corrected matter field. 

Once we have the evolved matter distribution, we apply the lensing data model and integrate along the line-of-sight with a ray tracer. We use the Born approximation and integrate radially from an observation point. From the three-dimensional density field, we compute the tidal field following Equation~\eqref{eq:tidal_field}. We then project the intrinsic alignments to the shear planes by integrating along the line-of-sight, Equation~\eqref{eq:projection_IA}. By combining the shear and intrinsic alignment fields, our method predicts the two components of the observed shear on the flat sky $\epsilon^b_{1,mn}$, $\epsilon^b_{2,mn}$ for each tomographic bin $b$ and $m,n$ sky pixel indices. 

The predicted shear will differ from the measured shear $\hat{\epsilon}^b_{1,mn}$, $\hat{\epsilon}^b_{2,mn}$ due to noise in the galaxy shape measurements. We account for that in the likelihood $P(\hat{\epsilon}^b_{1,mn}, \hat{\epsilon}^b_{2,mn}|\epsilon^b_{1,mn}, \epsilon^b_{2,mn})$. This method can handle shear noise that varies across the sky and tomographic bin. As in our previous work, we assume that the observations are characterised by shape noise with variance $\sigma_\epsilon^2$, and the associated shear uncertainty is given by the number $N_b$ of sources in a pixel in a tomographic bin, $\sigma_b = \sigma_\epsilon / \sqrt{N_b}$, where $\sigma_\epsilon=0.3$. The voxel likelihood can be approximated to be Gaussian with a variance $\sigma_b$ if $N_b$ is sufficiently large. Therefore, our log-likelihood is
\begin{align}
    \log \mathcal{L} &= \sum_b \sum_{mn} \log \left[ P(\hat{\epsilon}^b_{1,mn}, \hat{\epsilon}^b_{2,mn}|\epsilon^b_{1,mn}, \epsilon^b_{2,mn}) \right] \\
    & = - \frac{1}{2} \sum_b \sum_{mn} \frac{(\epsilon_{1,mn}^b - \hat{\epsilon}^b_{1,mn})^2 + (\epsilon_{2,mn}^b - \hat{\epsilon}^b_{2,mn})^2}{\sigma_b^2} + {\rm const.}, 
\end{align}
where the dependence of the observed shear on the initial conditions and underlying parameters is left implicit. 

In this work, we focus on two cosmological parameters ($\Omega_m$ and $\sigma_8$), assuming a flat Universe, $\Omega_\Lambda = 1 - \Omega_m$. We also sample the intrinsic alignment parameters $A_1$, $b_\mathrm{TA}$ and $A_2$, and the baryon feedback parameters $\beta$ and $\gamma$. We use the following uniform priors: $\Omega_\mathrm{m} \mathtt{\sim} {\cal U}[0.2, 0.7]$; $\sigma_8 \mathtt{\sim} {\cal U}[0.5,1.6]$; $A_1 \mathtt{\sim} {\cal U}[-5, 5]$; $b_\mathrm{TA} \mathtt{\sim} {\cal U}[-2, 2]$; $A_2 \mathtt{\sim} {\cal U}[-5,5]$; $\beta \mathtt{\sim} {\cal U}[0, 40]$; and $\gamma \mathtt{\sim} {\cal U}[1,2]$.

\subsection{Sampling scheme}
\label{sec:sampling}

We sample the posterior distribution, which requires varying the initial conditions, the cosmological parameters and the intrinsic alignment parameters.
Sampling the initial conditions $\delta^\mathrm{IC}$ implies that the density fluctuations in each voxel are a parameter of the problem. This results in a very high-dimensional space, and we used Hamiltonian Monte Carlo \citep{Neal2011} to deal with this high number of parameters. We used a slice sampler to sample the cosmological parameters $\boldsymbol{\theta}$. In this work, we have included another slice sampler for the TATT parameters $\boldsymbol{\chi}$ and one for the parameters of the EGD model $\boldsymbol{\xi}$. Sampling these additional parameters adds a minimal extra cost and allows us to propagate their uncertainties automatically. The different samplers are combined in a Gibbs sampling scheme, where the initial conditions, cosmology and nuisance parameters are sampled alternately as 
\begin{align}
    \boldsymbol{\delta}^\mathrm{ic} \curvearrowleft
    P(\boldsymbol{\delta}^\mathrm{ic}|\boldsymbol{\theta}, \boldsymbol{\chi}, \boldsymbol{\xi}, \boldsymbol{}\boldsymbol{d}), \\
    \boldsymbol{\theta} \curvearrowleft P(\boldsymbol{\theta} | \boldsymbol{\delta}^\mathrm{ic}, \boldsymbol{\chi}, \boldsymbol{\xi} , \boldsymbol{d}), \\
    \boldsymbol{\chi} \curvearrowleft P(\boldsymbol{\chi} | \boldsymbol{\delta}^\mathrm{ic}, \boldsymbol{\theta}, \boldsymbol{\xi} , \boldsymbol{d}), \\
    \boldsymbol{\xi} \curvearrowleft P(\boldsymbol{\xi} | \boldsymbol{\delta}^\mathrm{ic}, \boldsymbol{\theta}, \boldsymbol{\chi} , \boldsymbol{d}).
\end{align} 
This scheme allows us to have a very flexible model without incurring too much cost of development and tuning of the chain.

\section{Simulated datasets}
\label{sec:mock_data}

In this work, we use Lagrangian perturbation theory (LPT) as our model of gravitational clustering to compare the results to our previous work \citep{BORG-WL2022}. To test the effect of the intrinsic alignment model on the constraining power of the method, we use the same resolution and setup as in our earlier work. At that resolution, the baryon effects are negligible. We, therefore, also use a higher resolution and a smaller box to test and validate our implementation of the baryon feedback. This second box is too small for a cosmology analysis. We also used LPT to test the baryon feedback implementation. However, an accurate description of  the matter distribution at those scales would require a fully non-linear particle mesh.

We generate simulated data assuming a standard $\Lambda$CDM cosmology with $\Omega_m=0.3175$, $\Omega_\Lambda=0.6825$, $\Omega_b=0.049$, $h=0.677$, $\sigma_8=0.8$ and $n_s=0.9624$. Then, we generate initial conditions in a cartesian grid and evolve them via LPT, including light-cone effects. We used cloud-in-cell weighting to obtain the density field from the particles. We then generated shear fields with intrinsic alignments, following the forward model described in Section~\ref{sec:data_model}. The intrinsic alignment parameters are $A_1=0.18$, $b_\mathrm{TA}=0.8$ and $\ A_2=0.1$ \citep{DesRobustness}. We generate two datasets at different resolutions:

\begin{enumerate}
    \item Dataset 1: we used a similar setup as in our previous work \citep{BORG-WL2022} to investigate the effect of intrinsic alignments on the cosmology constraints. We used a box of $(1\times 1 \times 4.5) h^{-1}$~Gpc, with $64 \times 64 \times 128$ voxels, and the redshift distribution of sources shown in Figure~\ref{fig:tomobin}. This corresponds to an area of (16 deg)$^2$ with a resolution of 15 arcmins. We then added Gaussian pixel noise with a variance corresponding to 30 galaxies per square arcmin, uniformly distributed between the tomographic bins, as expected for Euclid \citep{Euclid2020CosmicShear}.  We use an uncertainty on intrinsic ellipticity of $\sigma_\epsilon = 0.3$, being this the variance of both shear components. At the resolution of this dataset, the baryon effects are negligible.
    \item Dataset 2: We generated a higher-resolution set of mock data to test the baryon feedback sampler and illustrate the effects of model misspecification for the baryons. We used a box of $(0.1 \times 0.1 \times 2.5 ) h^{-1}$~Gpc, with $64 \times 64 \times 128$ voxels, and the tomographic bins 1 and 2 shown in Figure~\ref{fig:tomobin}. This corresponds to an area of (3 deg)$^2$ with a resolution of 5~arcmins. Since we use this dataset to validate our baryon feedback implementation rather than testing the constraining power of the method, we use an unrealistic higher source density, six times higher than expected for upcoming surveys to have a $\mathrm{S/N}\approx1$ at the mean $\ell$ of the box. The tests with Dataset 2 are,  therefore, only for implementation testing and illustrative purposes.
\end{enumerate}

\begin{figure*}
    \includegraphics[width=2\columnwidth]{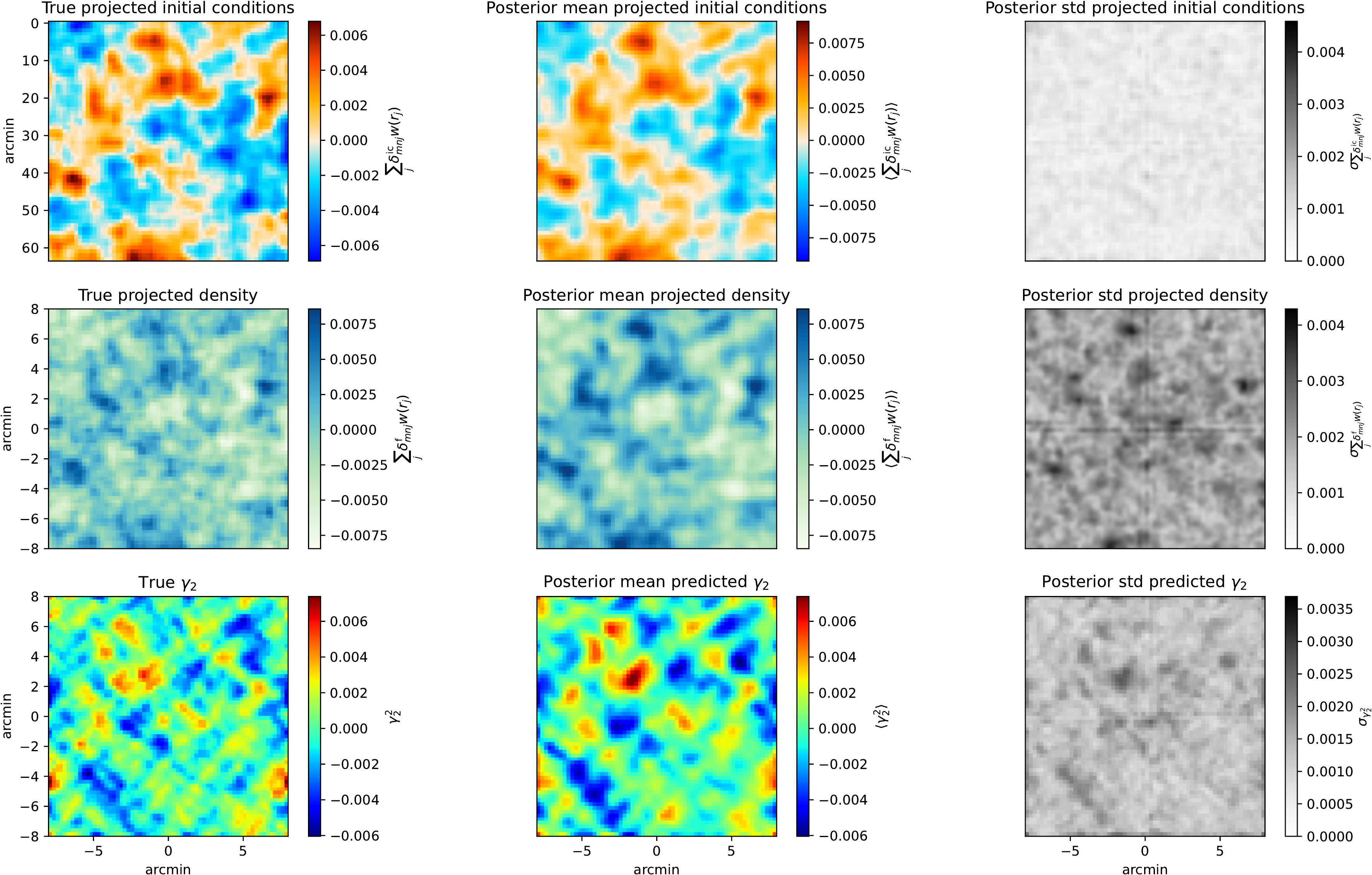}
    \caption{Projection of the initial (top row), evolved density fields (middle row)  and the posterior predicted shear fields (bottom row). The first column shows the true fields we used to generate the data, the second column shows the ensemble mean, and the third column shows the standard deviation of the fields. The mean and standard deviation are estimated from 400 effective MCMC samples. The artefact in the standard deviation is due to the line-of-sight projector, which ensures that all the lines-of-sight have the same physical length, resulting in a different number of voxels, and therefore higher uncertainty for the lines-of-sight with fewer voxels. This artefact affects only the standard deviation, which is not used in the inference.}
    \label{fig:panels}
\end{figure*}

\begin{figure*}
    \includegraphics[width=2\columnwidth]{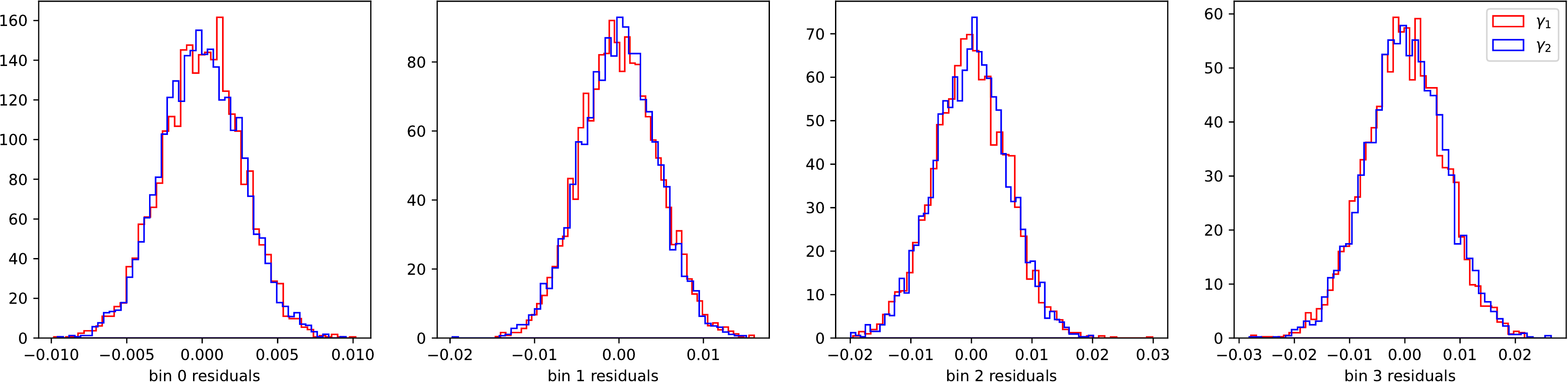}
    \caption{Residuals of the recovered shear fields.}
    \label{fig:residuals}
\end{figure*}

\section{Results}
\label{sec:results}

In this section, we detail the results in terms of mock constraints on inferred density field, cosmological parameters, and potential bias introduced by the TATT and EGP models. The validation tests of the method are described in Appendix~\ref{appendix:validation}.

\subsection{Cosmology and intrinsic alignments constraints}

Here we present the posterior constraints from applying BORG-WL to Dataset 1, described in Section~\ref{sec:mock_data}.

Figure~\ref{fig:TATT_contours} shows the posterior constraints on $\Omega_m$ and $\sigma_8$. To investigate how intrinsic alignments affect the constraining power of BORG-WL, we compared these with the constraints for another self-consistent analysis with an equivalent setup but setting $A_1=A_2=0$. Adding the TATT parameters weakens the constraints, but BORG-WL still lifts the weak lensing degeneracy and provides tight constraints on $\Omega_m$ and $\sigma_8$ from weak lensing alone. The size of the posterior distribution is significantly smaller than the priors for $\Omega_m$ and $\sigma_8$. Therefore, the constraints on these parameters should not be affected by the prior choice. We have added the constraints from the angular power spectra ($C_\ell$) for the same setup (see Appendix~\ref{appendix:cls}), showing that BORG-WL lifts the weak lensing degeneracy with and without intrinsic alignments\footnote{The contours without intrinsic alignments differ from the ones in \cite{BORG-WL2022} because we have changed the noise level and extended the size of the box in the radial direction to prevent the sources exiting the volume, which led to a slight truncation of the contours at $\Omega_m=0.2$ in Figure~3 of \cite{BORG-WL2022}.}.

Figure~\ref{fig:corner_TATT} shows the joint and marginal posterior distributions of the cosmological and TATT parameters. BORG-WL can constrain $A_1$ and $A_2$, but the lensing data at this resolution is not sufficiently informative to constrain $b_\mathrm{TA}$, and the marginal distribution of this parameter corresponds closely to the prior. The other distributions are significantly more compact than the size of the prior distributions. These posteriors are computed from 35~000 effective samples after the burn-in phase and are dominated by the data via the likelihood.

\begin{figure*}
    \includegraphics[width=0.9\hsize]{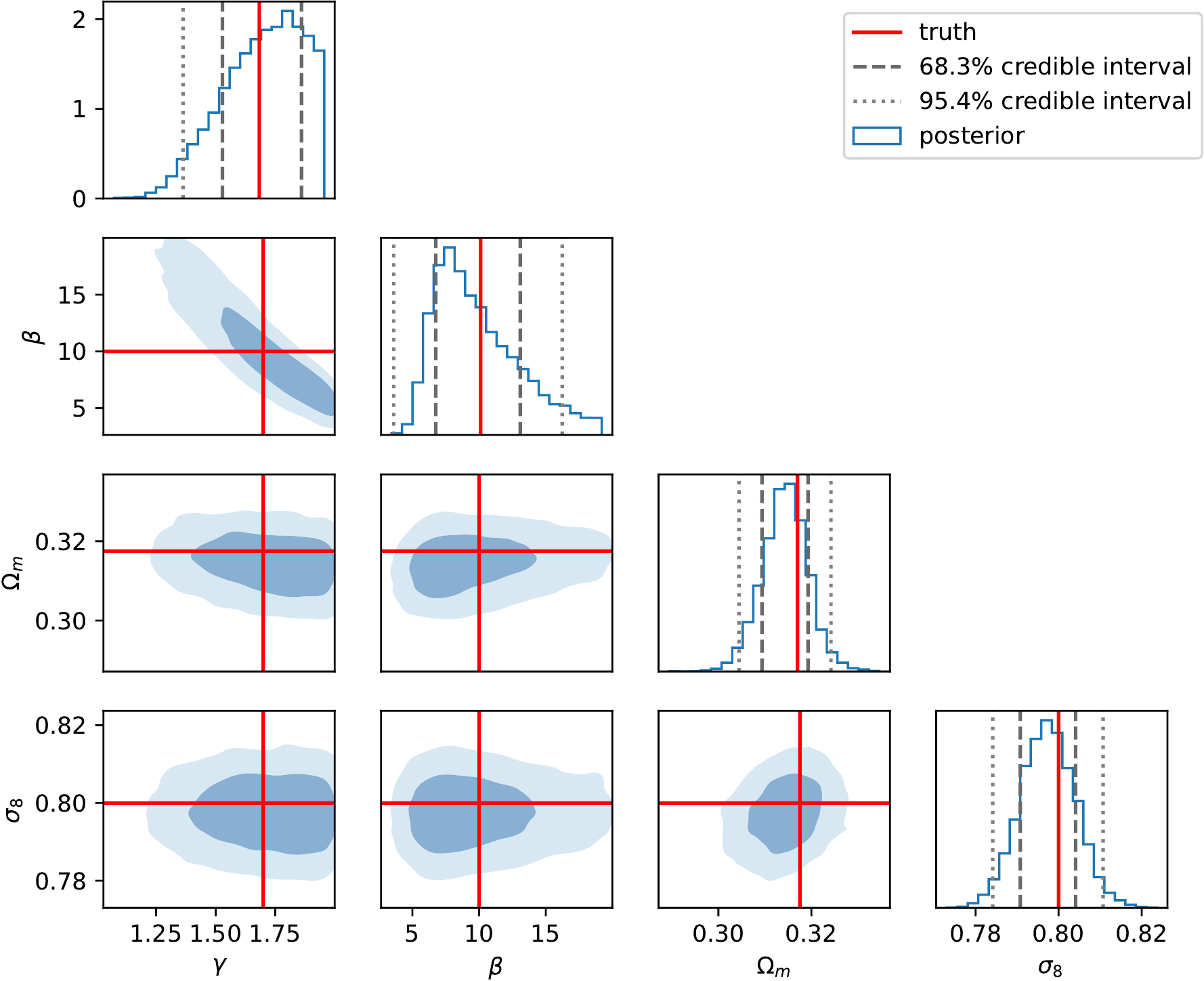}
    \caption{Posterior distribution of the cosmology and baryon parameters inferred from Dataset 2. The red lines indicate the truth, the grey dashed lines correspond to 68.3\% credible interval, and the dotted lines indicate the 95.4\% credible interval. All the posterior distributions are narrower than the size of the prior except  for $\gamma$, which has a uniform prior between $[1,2]$.}
    \label{fig:corner_baryons}
\end{figure*}

\begin{figure*}
    \includegraphics[width=0.9\hsize]{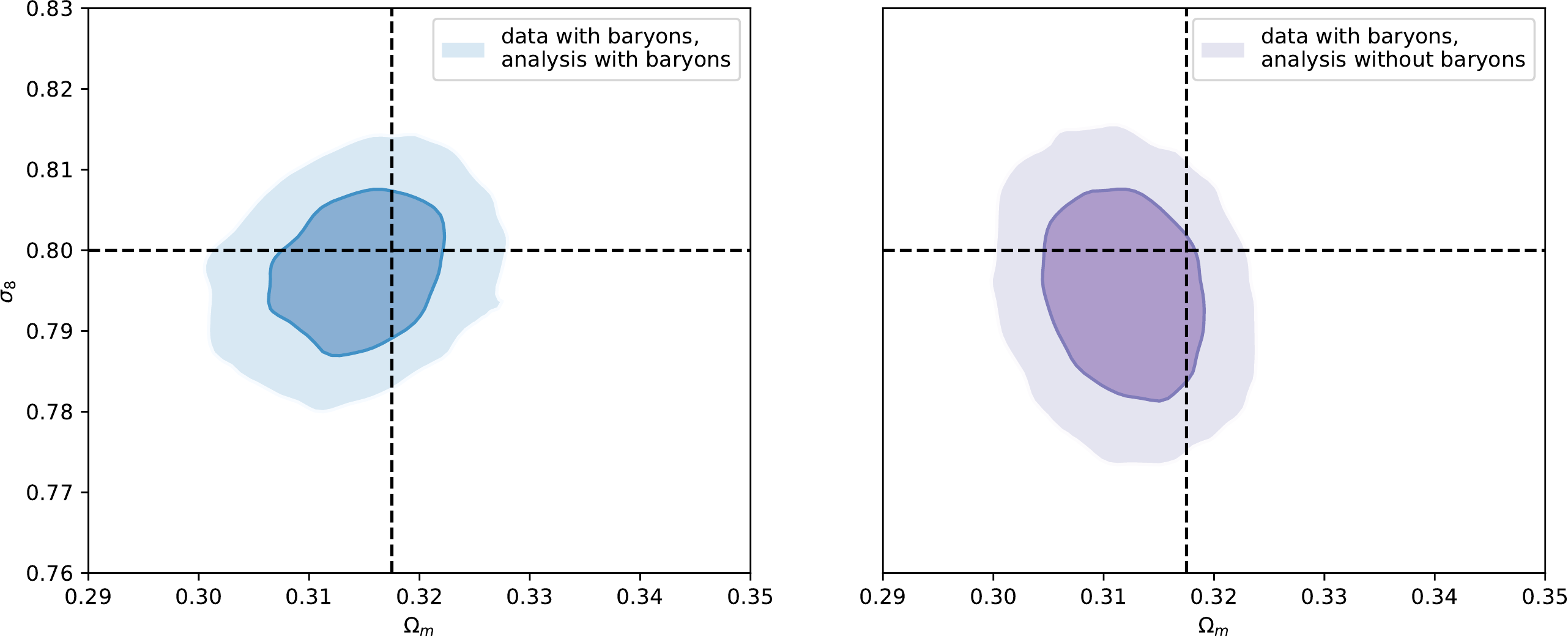}
    \caption{68.3\% and 95.4\%  highest posterior density credible regions from analysing Dataset 2 with the EGD model (left panel) and fixing the amplitude of the baryon displacement field to $\beta=0$ (right panel). The dashed lines indicate the true values of the parameters.}
    \label{fig:baryons_models}
\end{figure*}

\subsection{Model misspecification}

Here we study the effect of using a simpler model to describe the intrinsic alignments. Since the NLA model \citep{Bridle2007}
is also commonly used in weak lensing analyses \citep{Troxel2018, Asgari2021, Hikage2019, Hamana2020}, here we compare the cosmology posteriors of analysing Dataset 1 with the NLA model. We remind the reader that the NLA model is defined as
\begin{align}
    \gamma^\mathrm{IA}_1 (\boldsymbol{r}, \boldsymbol{\vartheta}) &= C_1 (s_{xx} - s_{yy}) \\
    \gamma^\mathrm{IA}_2  (\boldsymbol{r}, \boldsymbol{\vartheta}) &= 2 C_1 s_{xy}\;,
\end{align}
following the notations of Section~\ref{sec:ia_model}.  As opposed to the results discussed in the previous section, this is not a self-consistent test because the synthetic data is generated with the TATT model. We also note that the NLA model is equivalent to the TATT model in the limit that $C_{1\delta}=0$ and $C_2=0$.

Figure~\ref{fig:IA_models} shows the results of analysing the same synthetic data with two different options in the analysis pipeline: with the TATT model (self-consistent test) and with the NLA model (model misspecification). We find that the misspecification of the intrinsic alignment model weakens the constraints but does not bias the results at the resolution of this experiment. We included the equivalent constraints from the angular power spectra, showing that the field-level results are consistent with the results from the angular power spectrum ($C_\ell$). 

\subsection{Inferred density fields}

Jointly with the cosmological and intrinsic-alignment parameters, BORG-WL also infers the three-dimensional matter density field. Here we focus on validating the results at the field level for the self-consistent test with the TATT model. We have also presented this test in our previous works \citep{BORG-WL, BORG-WL2022}, but in this case, we have extended the physics model with the intrinsic alignments, and we sample the TATT parameters.  

We draw samples of the primordial matter fluctuations and the matter distribution from the posterior distribution. Figure~\ref{fig:panels} shows the sky-projection of the true fields and the corresponding mean and variance of the samples, computed from 400 effective samples. A visual comparison shows that we recover the structures from the truth. The mean fields show a smoothing effect, which is expected from averaging several samples. Figure~\ref{fig:residuals} shows the residual distribution for each tomographic bin.

\subsection{Baryon test}

To test our implementation of the baryon feedback and illustrate the effects of baryons on the cosmological constraints, we used Dataset 2 described in Section~\ref{sec:mock_data}. We also use LPT to describe structure formation for testing purposes in this analysis.  However, this model is not accurate at the small scales where the effect of baryons is relevant \citep{Tassev2013}.

We have analysed Dataset 2 sampling the baryon parameters $\gamma$ and $\beta$ jointly to the cosmological parameters $\Omega_m$ and $\sigma_8$ and the initial conditions. Figure~\ref{fig:corner_baryons} shows that our method can recover the correct values of the baryon and cosmological parameters. To test the effect of ignoring baryons, we have also analysed the same Dataset 2 imposing $\beta=0$ in the analysis pipeline. Figure~\ref{fig:baryons_models} compares the two analyses, showing that analysing the data without the baryon model can bias the constraints towards lower values of $\Omega_m$.

\section{Summary and conclusions}
\label{sec:conclusions}

We have presented an extension of our field-level approach to infer cosmological parameters and the dark matter distribution from weak lensing data \citep{BORG-WL,BORG-WL2022}. We have included intrinsic alignments and baryon feedback and sampled the parameters associated with these models. As a result, we have a forward model that includes sufficient physics for application to data.

We have added the TATT model \citep{Blazek2011} (and NLA as a subset) to our framework to describe the intrinsic alignments. Since BORG-WL infers the three-dimensional dark matter distribution, we can access the tidal field from which we compute the intrinsic alignments. We sample the TATT parameters jointly with the cosmology and the initial conditions, showing that BORG-WL recovers the correct values of the parameters without significantly reducing the constraining power of the method. We have also tested the effects of model misspecification.

We have also included the EGD model \citep{Dai2018} to account for baryon feedback and correct the density at the field level. We sample the parameters of the model, allowing it to self-calibrate from the data. We have illustrated the effects of the baryon feedback in a small volume of high-resolution synthetic data and shown that our method recovers the true values of the EGD parameters. 

After these extensions, the forward model consists of uniform priors for the cosmological, baryon and intrinsic alignment parameters; a Gaussian prior for the primordial fluctuations and a physical description of gravity and structure formation that accounts for baryon effects and links the initial conditions to the total matter density. The cosmological parameters are sampled, changing the matter power spectrum, geometry, tidal field, structure growth, and distance-redshift relation. This allows us to constrain the cosmology, matter density and systematics parameters simultaneously.

With this work, we have brought BORG-WL closer to the real data application to constrain cosmological parameters and the underlying dark matter distribution. Future work will focus on accounting for the uncertainty in the redshift distribution of sources and applying BORG-WL to real cosmic shear measurements.

\section*{Acknowledgements}
We thank David Alonso and Supranta Sarma Boruah for useful discussions. We thank  Eleni Tsaprazi for comments on the original manuscript. NP is supported by the Beecroft Trust. This work was partly supported by STFC through Imperial College Astrophysics Consolidated Grant ST/5000372/1, and partly supported by the Simons Collaboration on ``Learning the Universe''.  This work was granted access to the HPC resources of  TGCC (Irene Rome) under the allocation AD010413589. GL acknowledge support from the Centre National d’Etudes Spatiales, through the grant GCEuclidNext. This work was carried out within the Aquila Consortium\footnote{\url{https://www.aquila-consortium.org}}.

\section*{Data Availability}
 The data underlying this article will be shared on reasonable request.



\bibliographystyle{mnras}
\bibliography{lensing} 

\begin{figure*}
    \includegraphics[width=\hsize]{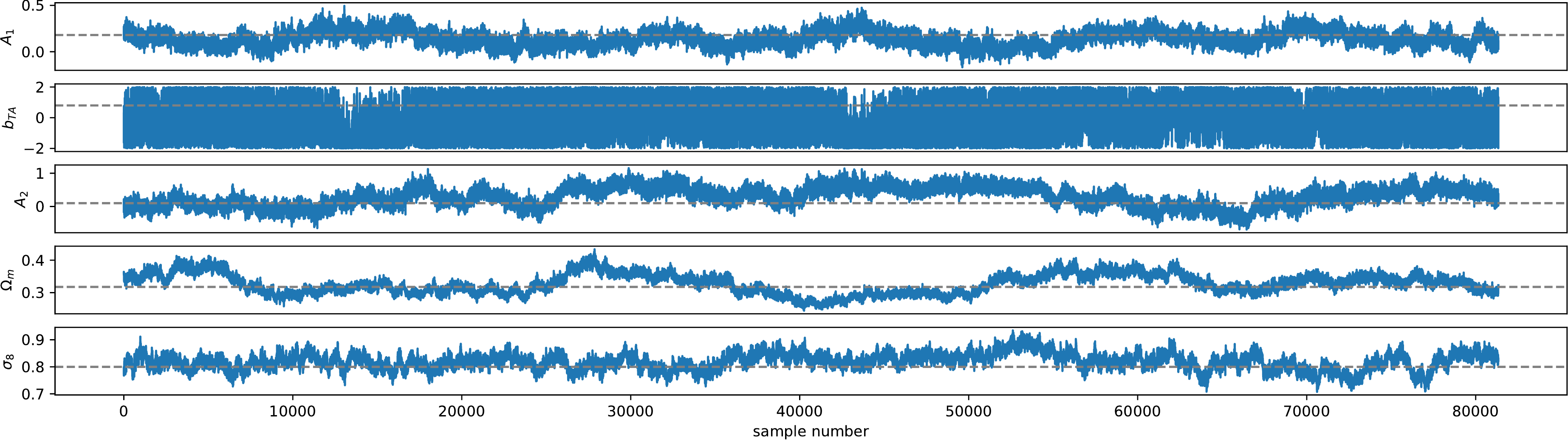}
    \caption{Trace plot of cosmological and TATT parameters after discarding the burn-in phase. The dashed lines indicate the true values.}
    \label{fig:trace}
\end{figure*}

\begin{figure}
    \includegraphics[width=\columnwidth]{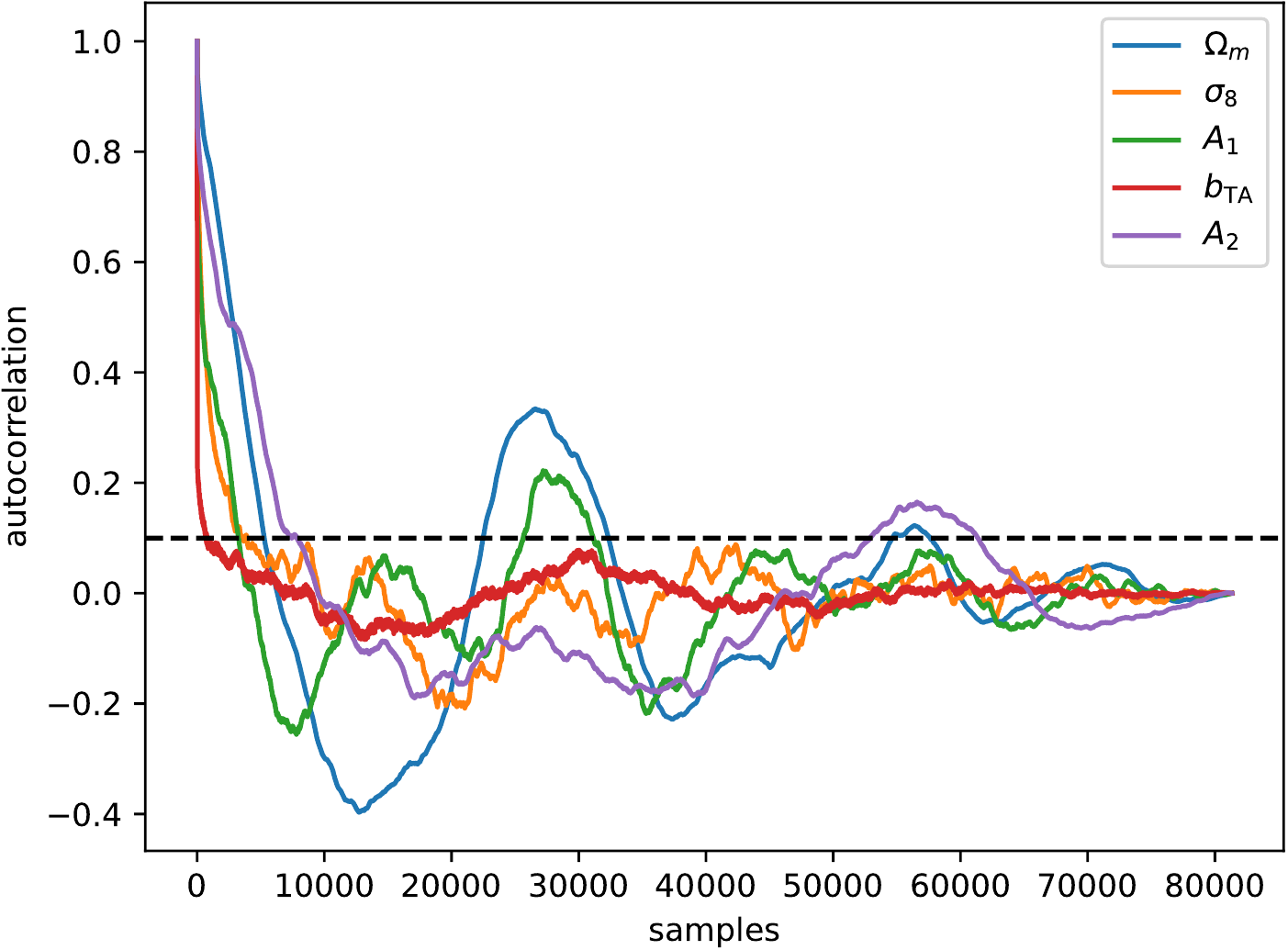}
    \caption{Auto-correlation of the cosmological parameters as a function of the sample number in the Markov chain. The correlation length of the sampler can be estimated as the point when the correlation drops below 0.1 (dashed line).}
    \label{fig:corr}
\end{figure}

\appendix

\section{Validation tests}
\label{appendix:validation}
In this appendix, we describe the validation tests of our method. 

Figure~\ref{fig:trace} shows the trace plots of the 50~000 samples in this analysis after the burn-in phase. These samples are used to compute the joint and marginal distributions shown in Figure~\ref{fig:corner_TATT}. 

Figure~\ref{fig:corr} shows the autocorrelation of the samples, showing that the correlation length is of the order of 70~000 samples, corresponding to the first time the autocorrelation falls below 0.1. Note that the correlation length of $\Omega_m$ can be reduced by a factor of 4 rotating the parameter space and sampling $S_8$ instead of $\sigma_8$ \citep{BORG-WL2022}. 

We have assessed the convergence of the Markov chain with the \cite{Gelman92} test, which compares the variances between multiple chains with different starting points. For this test, we initialise two chains with different cosmology and different initial conditions. The Gelman-Rubin diagnostic is $R<1.05$ for all the cosmological parameters, indicating that the chains are converged.

\section{Two-point statistics constraints}
\label{appendix:cls}

Here we describe the model we used to generate all the constraints from the angular power spectra in the paper. 
We generated mock data in the same $\ell$-range and $n(z)$ as the Dataset 1 described in Section~\ref{sec:mock_data} as
\begin{equation}
    \hat{C}_\ell^{ij} = C^{ij}(\ell) + N \delta_{ij}
\end{equation}
where the indices $ij$ label the tomographic bins, $C^{ij}(\ell)$ is computed with the Core Cosmology Library \citep{CCL}, $\delta_{ij}$ is a Kronecker delta and $N$ is the noise, given by
\begin{equation}
    N = \frac{\sigma_\epsilon^2}{2 N_\mathrm{gal}},
\end{equation}
with $\sigma_\epsilon = 0.3$ and $N_\mathrm{gal}$ being the density of galaxies, 30 sources/arcmin$^2$ equally distributed between the tomographic bins. We did not include noise in the cross-power spectra, ignoring the small overlap between the tomographic bins.

To analyse these angular power spectra, we use a Gaussian log-likelihood
\begin{eqnarray}
     \log \mathcal{L} &=& -\frac{1}{2} \left[\hat{C}_{ij} - (C_{ij} + N\delta_{ij}) \right]^T \Sigma^{-1} \left[\hat{C}_{ij} - (C_{ij} + N\delta_{ij}) \right] \nonumber \\ &-& \frac{1}{2} \log (|\Sigma|)
\end{eqnarray}
where $\Sigma$ is the covariance matrix, computed at the true cosmology, as 
\begin{equation}
    \Sigma\left(C_\ell^{ab}, C_{\ell'}^{cd}\right) = \frac{C_\ell^{ac} C_\ell^{bd} + C_\ell^{ad} C_\ell^{bc}}{ (2\ell+1) f_\mathrm{sky} \Delta \ell} \delta_{\ell \ell'}, 
\end{equation}
and $f_\mathrm{sky} = A / (4\pi)$ is the fraction of the sky area covered by the survey. To sample, we used the MCMC package \textsc{emcee} \citep{emcee, Goodman2010} with uniform priors $\Omega_\mathrm{m} \mathtt{\sim} {\cal U}[0.01, 1.0]$; $\sigma_8 \mathtt{\sim} {\cal U}[0.4, 1.5]$; $A_1 \mathtt{\sim} {\cal U}[-5, 5]$; $b_\mathrm{TA} \mathtt{\sim} {\cal U}[-2, 2]$ and $A_2 \mathtt{\sim} {\cal U}[-5,5]$. We sampled $\Omega_m$, $\sigma_8$ and the intrinsic alignment parameters.

\bsp	
\label{lastpage}
\end{document}